\documentclass[3p,times]{elsarticle}

\usepackage{ecrc}
\usepackage{epstopdf}
\volume{00}
\firstpage{1}
\journalname{Nuclear Physics A}
\runauth{C. S. Fischer et al.}
\jid{nupha}

\jnltitlelogo{Nuclear Physics A}

\usepackage{graphicx}
\usepackage{amsmath,amssymb}

\begin{document}

\begin{frontmatter}

\title{Locating the critical end point of QCD}

\author[label1]{Christian~S.~Fischer}
\author[label2,label3]{Jan~Luecker}
\author[label1]{Christian~A.~Welzbacher}

\address[label1]{Institut f\"ur Theoretische Physik, Justus-Liebig-Universit\"at Gie\ss{}en, Heinrich-Buff-Ring 16, D-35392 Gie\ss{}en, Germany.}
\address[label2]{Institut f\"ur Theoretische Physik, Universit\"{a}t Heidelberg, Philosophenweg 16, D-69120
Heidelberg, Germany.}\address[label3]{Institut f\"ur Theoretische Physik, Goethe-Universit\"at Frankfurt, Max-von-Laue-Stra\ss{}e 1, D-60438 Frankfurt/Main, Germany}

\begin{abstract}
We summarize recent results for the phase structure of QCD at finite temperature and 
light-quark chemical potential for $N_f=2+1$ and $N_f=2+1+1$ dynamical quark flavors. 
We discuss order parameters for the chiral and deconfinement transitions obtained from
solutions of a coupled set of (truncated) Dyson-Schwinger equations for the quark and gluon 
propagators of Landau gauge QCD. Based on excellent agreement with results from lattice-QCD
at zero chemical potential we study the possible appearance of a critical end-point at large 
chemical potential. 
\end{abstract}


\end{frontmatter}



\section{Introduction}
\label{intro}

Heavy ion collision experiments at BNL, LHC and the future FAIR facility are designed
to probe the properties of the quark-gluon plasma (QGP) and the transition from
(almost) chirally symmetric and deconfined matter into the chirally broken and confined 
states of the hadronic phase. At the very large energies available at the LHC,
in principle two entire quark families have to be taken into account in the theoretical
description of these experiments via the equation of state. However, even at smaller 
temperatures at or above the light-quark crossover region, the effects of charm quarks 
on the EoS and the transition temperatures may not be entirely negligible. First lattice 
studies for $N_f=2+1+1$ flavors indicate indeed, that charm quarks may not be treated in 
the quenched approximation, i.e. the back-reaction of the charm quarks onto the Yang-Mills 
sector of the theory is quantitatively important \cite{Ratti:2013uta}.

In this contribution we summarize recent results on the QCD phase diagram via a framework 
of Dyson-Schwinger equations (DSEs) for the quark and gluon propagator of Landau gauge QCD. 
We discuss the phase diagram for $N_f=2+1$ flavors and estimate the influence of the charm 
quark on the phase structure of QCD and the location of a putative critical end-point. 
While the approach is first principle, truncations have to be introduced to convert the 
equations into a form suitable for practical calculations. In order to make these
approximations well controlled we use constraints such as symmetries and conservation laws 
as well as comparison with corresponding results from lattice calculations (where available). 
Our aim is to tighten this control to such extent that reliable results for large chemical 
potential become feasible. One of the advantages of our framework over model treatments 
(see e.g. Ref~\cite{Qin:2010nq} and Refs. therein) is the direct accessibility of the 
Yang-Mills sector of QCD thus rendering a fully 
dynamical treatment of all members of the first two quark families feasible.
For corresponding efforts in the functional renormalization group approach see e.g.
\cite{Braun:2009gm,Braun:2007bx,Herbst:2010rf,Herbst:2013ail,Fister:2013bh,Herbst:2013ufa} 
and the overview of Jan Pawlowski in these proceedings \cite{PawlowskiQM14}. 

This contribution is organized as follows. In the next section we summarize our approach, 
details can be found in \cite{Fischer:2009wc,Fischer:2010fx,Fischer:2012vc,Fischer:2013eca,Fischer:2014ata}. 
In section \ref{sec:results} we discuss results for $N_f=2+1$ quark flavors 
and compare with lattice QCD. We also study the effects of the charm quarks 
onto the QCD phase diagram.

\section{Framework}

\begin{figure}[t]
\includegraphics[width=0.45\textwidth]{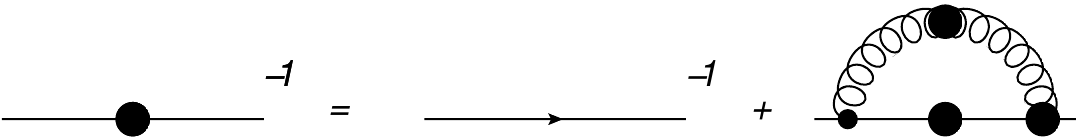}\hfill
\includegraphics[width=0.45\textwidth]{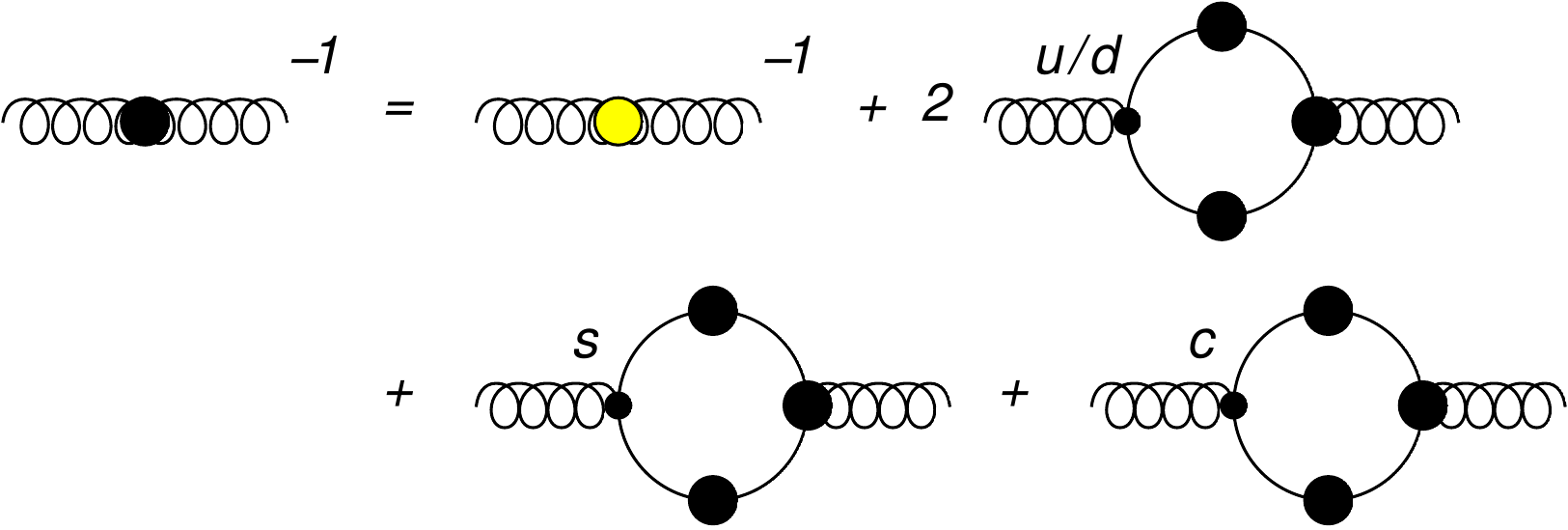}
\caption{Left: The DSE for the quark propagator. Large blobs denote dressed 
propagators and vertices. Right: The truncated gluon DSE for $N_f=2+1+1$ QCD. 
The light shaded (yellow) dot denotes the quenched (lattice) propagator. \label{fig:apprGluonDSE}}
\end{figure}

The DSEs for the quark and gluon propagators are displayed
in Fig.~\ref{fig:apprGluonDSE}. The fully dressed propagators on the left hand
side of the equations are determined from the diagrams on the right hand side,
which include the corresponding bare propagators as well as loops representing 
integrals in momentum space over non-perturbative propagators and vertices. In
the quark-DSE the essential ingredients are the fully dressed
gluon propagator as well as the dressed quark-gluon vertex. The appearance of the
dressed quark propagator also inside the loop renders the equation self-referential. 
We solve one quark-DSE iteratively for each of the $N_f$ quark flavors
involved in the calculation.

At the same time, each quark flavor is back-coupled onto the gluon propagator
according to the gluon-DSE displayed in the right part of Fig.~\ref{fig:apprGluonDSE}
for the case of $N_f=2+1+1$.
The key idea of our truncation of the gluon DSE is to replace the self-energy loops
containing only Yang-Mills propagators and vertices with the quenched gluon propagator 
obtained from lattice calculations. This procedure simplifies the numerical task 
dramatically and guarantees that the important physics of temperature effects in 
pure Yang-Mills theory is contained in our approximation scheme. In addition, we
use an ansatz for the quark-gluon vertex which contains temperature and chemical potential
effects according to (approximated) Slavnov-Taylor identities;
see Ref.~\cite{Fischer:2012vc} for details. This construction for the
vertex contains one open parameter, which controls the interaction strength of the
vertex at low momentum. This parameter can be fixed in two ways: either it is chosen
such that the pseudo-critical temperature of the chiral transition extracted from
$N_f=2+1$-lattice calculations is reproduced, or it can be fixed from experimental
input for the pion decay constant at zero temperature. The difference between these
two methods is a signal for the systematic error of our calculation and is at the
level of ten to fifteen percent for the pseudo-critical temperatures at zero 
chemical potential.  

\section{Results}\label{sec:results}

\begin{figure}[t]
{\includegraphics[width=0.38\linewidth]{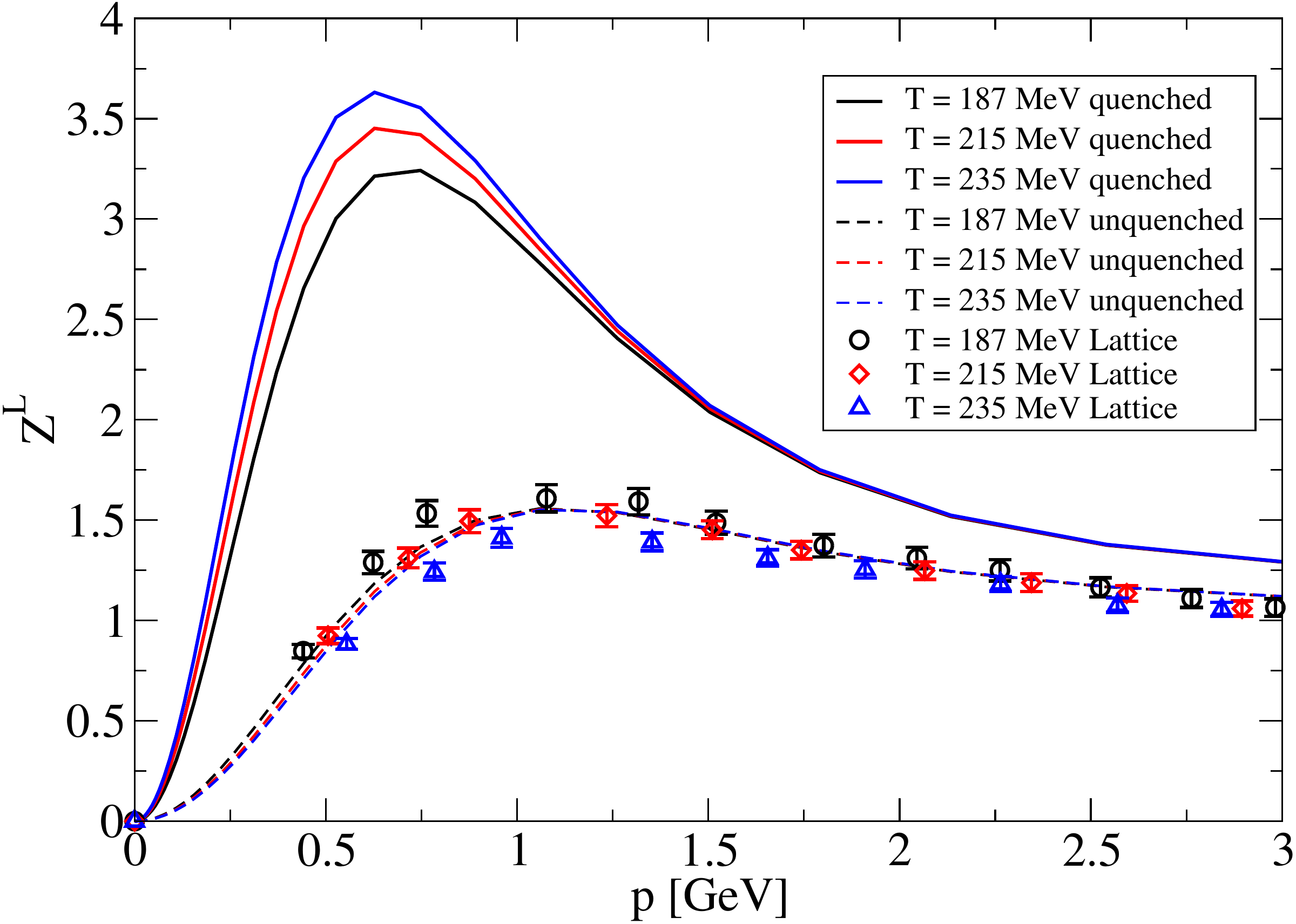}}\hfill
\includegraphics[width=0.39\textwidth]{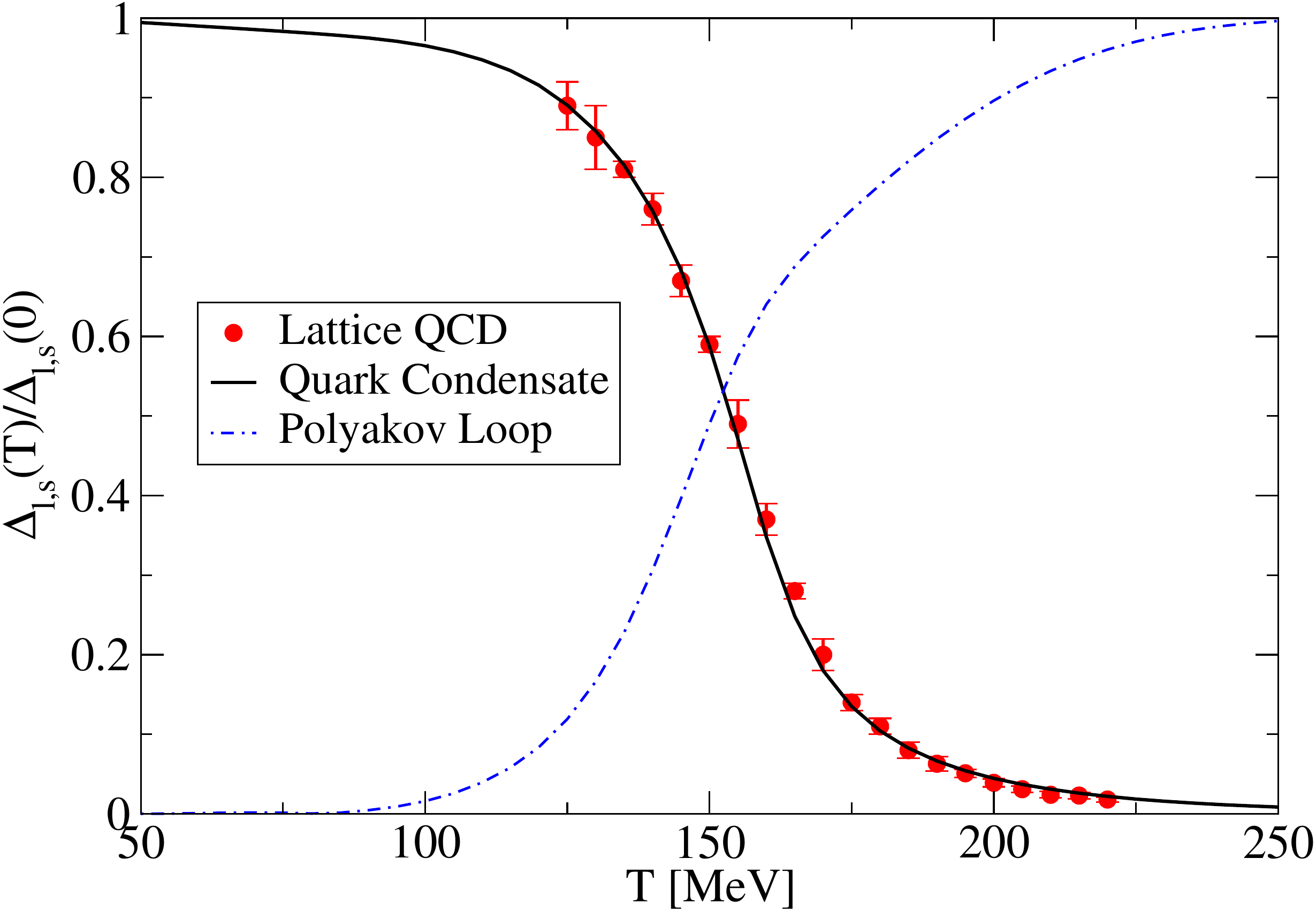}\hfill
\caption{Left diagram: comparison of electric gluon dressing function for $N_f$=2 in 
the DSE approach \cite{Fischer:2013eca} with lattice data \cite{Aouane:2012bk}. 
All results have been evaluated at a pion mass of $m_\pi = 316$ MeV. 
Right diagram: regularized chiral condensate and the Polyakov loop 
for $N_f=2+1$ quark flavors as a function of temperature at zero quark chemical 
potential $\mu_q = 0$. The interaction strength of the quark-gluon vertex is fixed to 
reproduce the critical temperature of the lattice results of Ref.~\cite{Borsanyi:2010bp}. 
Both figures are adapted from Ref.~\cite{Fischer:2014ata}. 
\label{fig:regcond}}
\includegraphics[width=0.38\textwidth]{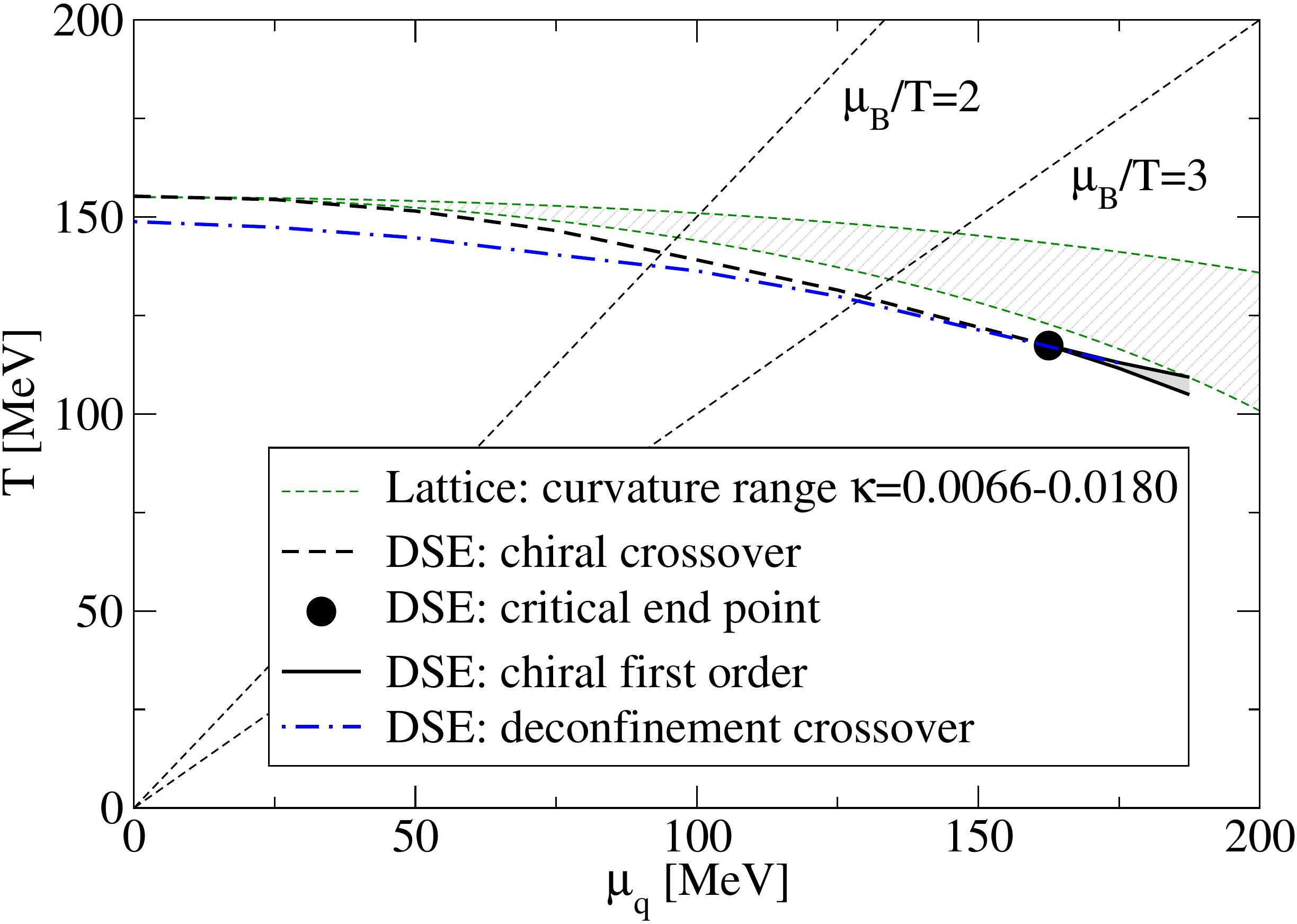}\hfill
\includegraphics[width=0.41\textwidth]{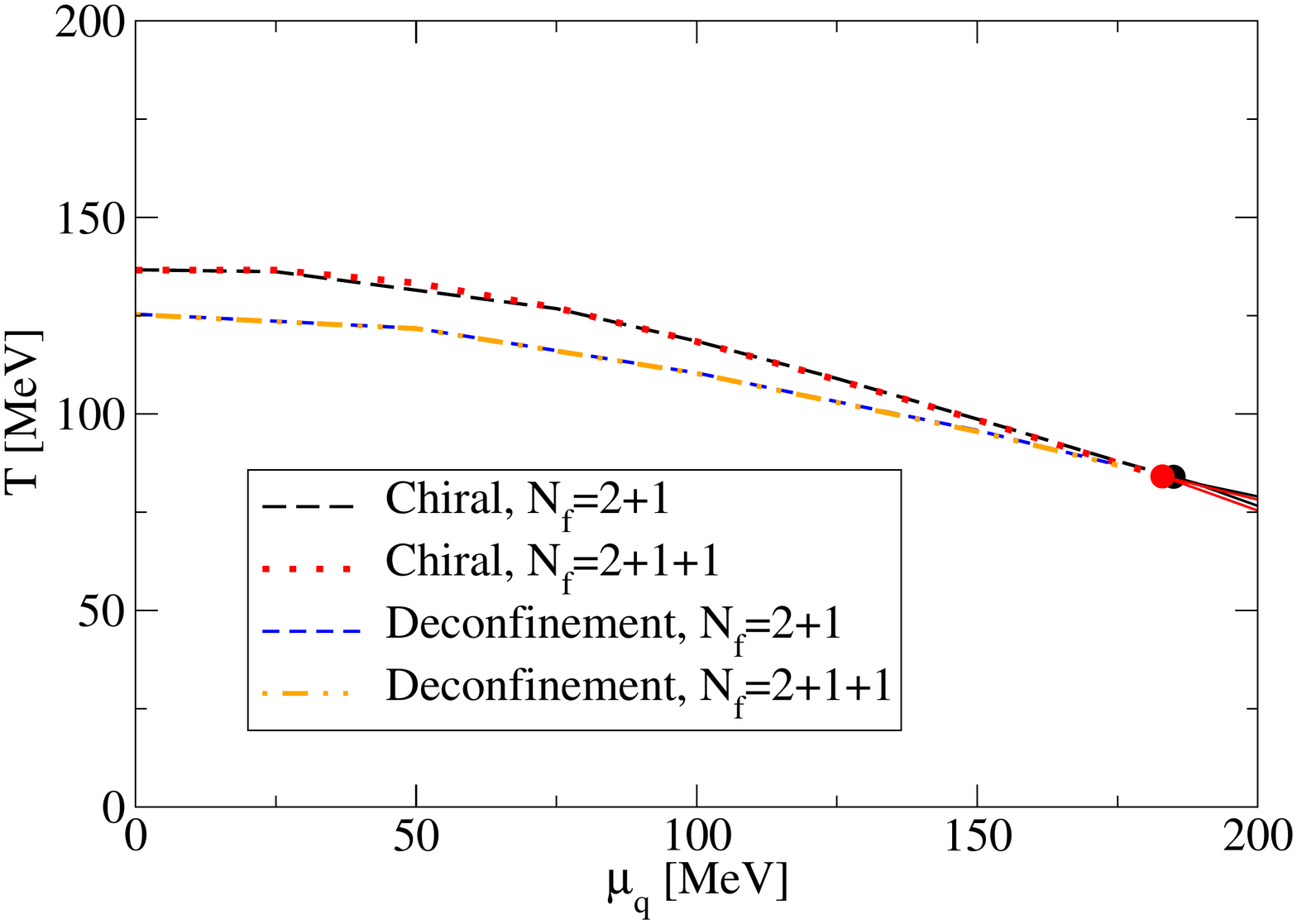}
\caption{Left diagram: phase diagram for $N_f=2+1$ quark flavors.
Shown are our results together with an extrapolation of a range of curvatures for 
the chiral transition extracted at imaginary and small chemical potential from 
different lattice groups \cite{Endrodi:2011gv,Kaczmarek:2011zz,Cea:2014xva}. 
Right diagram: comparison between $N_f=2+1$ and $N_f=2+1+1$ flavors, where in both cases 
the interaction strength of the quark-gluon vertex and the quark masses are fixed to 
reproduce vacuum physics. Both figures are adapted from Ref.~\cite{Fischer:2014ata}. 
\label{fig:phasediagram}}
\end{figure}

The quality of our approach may be assessed from the results for the unquenched
gluon propagator at finite temperature calculated in Ref.~\cite{Fischer:2012vc}. In 
Fig.~\ref{fig:regcond} we show the part of the propagator longitudinal to the heat bath
evaluated for $N_f=2$ and a heavy pion mass of $m_\pi=316$ MeV compared to 
corresponding lattice calculations of Ref.~\cite{Aouane:2012bk}. As can be seen from the
plot and is discussed in detail in Ref.~\cite{Fischer:2014ata}, both calculations match very 
well. We wish to emphasize that the DSE results appeared before the lattice data
and therefore constituted a prediction for the unquenched gluon.

Next we discuss the results for the chiral and deconfinement transition with $N_f=2+1$ 
physical up/down and strange quark masses \cite{Fischer:2014ata}. 
In the right diagram of Fig.~\ref{fig:regcond} we display the regularized quark 
condensate as well as the Polyakov loop as a function of temperature at zero chemical 
potential. We find excellent agreement with the lattice data. Here the strength of the 
quark-gluon interaction has been adjusted such that the (pseudo)-critical
transition temperature of the lattice data is reproduced. A non-trivial result, however,
is the agreement of the slope of the chiral transition of the DSE results with the lattice data
(see Ref.~\cite{Herbst:2013ufa} for a corresponding agreement within the PQM model). Together with 
the results for the unquenched gluon this demonstrates that out truncation scheme works well
at zero chemical potential. The resulting transition temperatures extracted from the chiral 
susceptibility and the inflection point of the light quark condensate are given by 
\begin{equation}
\left.T_c\right|_{\frac{d\langle\bar\psi\psi\rangle}{dm}} = 160.2 \,\mbox{MeV}\,, \hspace*{2cm} 
\left.T_c\right|_{\frac{d\langle\bar\psi\psi\rangle}{dT}} = 155.6 \,\mbox{MeV}.
\end{equation}

Our results for the QCD phase diagram at finite chemical quark potential are shown
in the left diagram of Fig.~\ref{fig:phasediagram} using the inflection point of the light 
quark condensate to pin down 
the transition temperature for the chiral transition and the Polyakov loop potential
for the corresponding one for deconfinement. The chiral crossover, displayed by the 
dashed black line turns into a chiral critical end point (CEP) at 
\begin{equation}
(T^c,\mu_q^c)=(115,168) \,\mbox{MeV}, 
\end{equation}
where also the deconfinement transition line meets the chiral one. To better guide 
the eye, we have also plotted lines with ratios of baryon chemical potential over 
temperature $\mu_B/T = 2$ and $\mu_B/T = 3$.

In order to assess the quantitative reliability of our result for the CEP some comments are
in order. In the current approximation for the quark-gluon vertex we relied on a construction
that is built along its Slavnov-Taylor identity, but does not yet encode explicitly the effects 
of the back-coupling of mesons and baryons onto the quark propagator (see e.g.
Refs.~\cite{Fischer:2007ze,Fischer:2008wy,Fischer:2011pk} for first studies of these effects).
At $\mu = 0$ and finite quark mass it appears not to be necessary to include these effects 
explicitly, as demonstrated by the agreement with the quark condensate from the lattice. However,
finite chemical potential, meson and baryon effects in the quark-gluon vertex may introduce 
additional dependence on chemical potential on top of the one already covered by our ansatz.
It certainly needs to be studied in future work, whether these contributions have a large
impact onto the location of the critical end point.

In Fig.~\ref{fig:phasediagram} we also compare our result for the chiral transition line with 
lattice results of its curvature determined around $\mu=0$ and extrapolated to
finite chemical potential \cite{Endrodi:2011gv,Kaczmarek:2011zz,Cea:2014xva}.
Although the overall agreement of the lattice extrapolation with our results is quite
satisfactory one should bear in mind the above caveats. Similar caveats may apply to
the lattice extrapolations, since the effects of baryons onto the curvature may only
set in at large chemical potential and may therefore not be captured by the extrapolation.

The effects of charm quarks onto the chiral transition line can be assessed from the diagram
on the right hand side of Fig.~\ref{fig:phasediagram}. To this end we fixed both, the interaction
strength of the quark-gluon vertex for $N_f=2+1$ \emph{and} $N_f=2+1+1$ to the physical
pion decay constant. As mentioned above, this leads to a decrease of the chiral transition
temperature by about $\Delta T \approx 20$ MeV compared to the results discussed above. 
On the other hand, this procedure taking care of the change of scale when going from 
$N_f=2+1$ to $N_f=2+1+1$ allows us to directly compare the phase diagram with and without
charm quarks. As a result we find that the influence of the charm quark on the chiral and 
deconfinement transition is almost negligible apart from a very small shift of the critical 
end point towards smaller chemical potential still within the limits of numerical 
uncertainty\footnote{If instead we kept the interaction strength and scale of the $N_f=2+1$
case and merely added the charm quark we would have obtained a shift of the transition
line towards smaller temperatures of about $\Delta T \approx 23$ MeV.}. 
We expect to see a similar behavior in corresponding future lattice calculations.

\vspace*{0mm}
{\bf Acknowledgments}\\
This work has been supported by the Helmholtz International Center for 
FAIR within the LOEWE program of the State of Hesse.


\end{document}